\documentstyle[preprint,aps]{revtex}
\begin{document}
\draft
\tightenlines

\title{Ground state fluctuations in finite Fermi and Bose systems}
\author{Muoi Tran, M. V. N. Murthy\footnote{Permanant Address; The 
Institute of Mathematical Sciences, CIT Campus, Chennai 600 113, India},
and R. K. Bhaduri}

\address
{Department of Physics and Astronomy, McMaster University\\
 Hamilton, Ont. L8S 4M1, Canada}
\date{\today}
\maketitle

\begin{abstract} 
We consider a small and fixed number of fermions (bosons) in a trap. 
The ground state of the system is defined at $T=0$. For a 
given excitation energy, there are several ways of exciting the particles  
from this ground state. We formulate a method for calculating the 
number fluctuation in the ground state using microcanonical counting, and 
implement it for small systems of noninteracting fermions as well as
bosons in harmonic confinement. This exact calculation for
fluctuation, when compared with canonical ensemble averaging, gives 
considerably different results, specially for fermions. 
This difference is expected to persist at low excitation even when the 
fermion number in the trap is large.

\end{abstract}


\narrowtext 
\section{Introduction}

The traditional approach to determining the number fluctuations in a 
given quantum state relies on the statistical description of the 
system based on the grand canonical ensemble (GCE). For bosons, if 
$<N_i>$ is the average occupancy of a single orbital $i$, then the 
mean-square fluctuation in GCE is given by~\cite{kittel} 
$<(\Delta N_i)^2>=<N_i>(1+<N_i>)$. As the temperature 
$T \rightarrow 0$ for the boson system, however, there is 
macroscopic occupancy in 
the ground state, with the result that the relative fluctuation 
$ <(\Delta N_0)^2>^{1/2}/N \rightarrow 1$. This is clearly 
absurd, since at $T=0$, none of the bosons 
(in the noninteracting case) should be in the excited orbitals, so the fluctuation should vanish.   
This problem, first discussed by Fujiwara et al.~\cite{fujiwara}, is 
of topical interest with the discovery of Bose-Einstein condensation 
in a dilute gas of alkali atoms confined in a trap
~\cite{experiments}. The problem of number fluctuation from the 
ground state of a system of confined fermions 
is also of interest due to the recent
developments in the field of quantum dots\cite{dotexpt}.  
The fluctuation
in Bose-Einstein systems has been addressed in a series of
papers\cite{grossman1,grossman2,grossman3,gajda,illuminati}
both in micro-canonical and canonical ensemble framework. A clear and
unambiguous definition of the fluctuation has been given for 
calculating the ground state occupation number fluctuation by Grossman and
Holthaus\cite{grossman2} through the "microcanonical entropy" of an
N-particle Bose system. Detailed comparisons between the two methods 
for noninteracting bosons in a harmonic trap have been made by Holthaus and 
Kalinowski~\cite{grossman3}, where analytical expressions for the two 
are also given using a modified saddle-point approximation.   
   
Our focus in this paper is on the exact calculation of the ground state 
fluctuation for a small number of fermions confined in a trap. For comparison, 
the corresponding calculations are also done for bosons.  
To this end, a combinatorial method based on microcanonical counting 
is developed in sect.(II), and calculations are made for particles in one and 
two-dimensional harmonic traps. These exact results are then compared with the 
corresponding canonical ensemble calculations. The two methods yield 
substantially different, though qualitatively similar results. 
As explained later, we expect this deficiency of the canonical
ensemble averaging method to persist for fermions at low excitation,
even when the fermion number in the trap is large.
In sect.(III) we show that even though the canonical entropies for 
noninteracting bosons and fermions in a one-dimensional harmonic trap
are identical, the number fluctuations in the ground state are vastly 
different. The numerical results are discussed in sect.(IV).

\section{Fluctuations in the independent particle model}

We assume an independent particle model where the one particle 
partition function is given analytically, or may be computed. We 
first give the well-known recipe of calculating the $N$- particle 
quantum canonical partition system from this. At zero temperature all 
the particles are in the ground state. At a finite temperature, 
however, the excitation energy may be shared in many different ways 
amongst the particles, with the result that the population in the 
original ground state is not fixed, although the total number, $N$, 
is still the same. Our objective here is to define and calculate this
fluctuation in the ground state occupation as a function of excitation
energy or, equivalently, temperature. 

The canonical partition function for bosons and fermions in any space
dimension may be written as 
\begin{equation}
Z_N^{B,F} = (\pm)^N \sum_{n_1,n_2,...,n_N} 
\prod_{j=1}^{N}\frac{[\pm Z_1(j\beta)/j]^{n_j}}{n_j!},
\label{master}
\end{equation}
where $Z_1(\beta)$ is the single particle partition function, $\beta$ is
the inverse temperature and the upper and lower signs refer to bosons and
fermions respectively.  The sum over the set of integers $n_i$ is
constrained by the relation 
\begin{equation}
\sum_{j=1}^N~j n_j=N~.
\label{constraint}
\end{equation}
The above formula allows us to write a general recursion relationship for
the canonical partition function 
\begin{equation}
Z_N^{B,F} = \frac{1}{N} \sum_{n=1}^N (\pm)^{n+1} Z_1(n\beta)
Z_{N-n}^{B,F}(\beta)
\label{recursion} 
\end{equation}
for bosons $(+)$ and fermions $(-)$. We note that in the above recursion 
relation $Z_0$ is formally taken to be unity for consistency.

In order to perform explicit calculations, we specialise to the case of a 
harmonic oscillator in d-dimensions. The single particle partition function is
given by,
\begin{equation}
Z_1(\beta) = [\frac{x^{1/2}}{(1-x)}]^d
\end{equation}
where 
\begin{equation}
x = exp(-\beta\hbar\omega).
\label{kala}
\end{equation}

The canonical partition function for a system with $N$ particles is then
computed using Eq.(\ref{master}) and is given by
\begin{equation}
Z_N = x^{Nd/2} P_N(x) \prod_{j=1}^N \frac{1}{(1-x^j)^d},
\end{equation}
where $P_N(x)$ is a polynomial in x 
which depends on the dimension and the
statistics of the system. We shall use the notation $P_N(x)=B_N(x)$ and
$P_N(x)=F_N(x)$  for bosons and Fermions where necessary.
The polynomial may be calculated using the recursion relation in
Eq.(\ref{recursion})
\begin{equation}
P_N(x) = \frac{1}{N} \sum_{n=1}^N (\pm)^{n+1}\frac{ \prod_{j=N-n+1}^{N}
(1-x^j)^d }{(1-x^n)^d} P_{N-n}(x). 
\end{equation}
The recursion relation above should be used with  the condition $P_0(x)=1$
for both bosons and fermions. We further note that in one-dimension
$B_N(x)=1$ for bosons and $F_N(x)=x^{N(N-1)/2}$ for fermions. They are
however more complicated in higher dimensions~\cite{schmidt}.

\subsection{Fluctuations from Microcanonical Counting}
We first define the fluctuation in particle number 
from the
ground state at a given excitation energy through a set of counting 
rules. Again we first write down the general formulae for a given a set of
discrete energy levels, and then specialize to the harmonic trap.
The single particle partition function may be 
written as
\begin{equation}
Z_1(\beta) = x^{\epsilon_0}\sum_{j=1}^{\infty} x^{\epsilon_j},
\end{equation}
where $ x=e^{-\beta} $ and $\epsilon_j , j=0,..,\infty$ are the single
particle energies. It is understood that $\beta$ in the exponent defining 
$x$ has been multiplied by a characteristic energy scale of the system, 
and similarly $\epsilon_j$ has been divided by the same, which we put to 
unity for convenience. For the harmonic oscillator, this energy scale is 
$\hbar\omega$, and $x$ is given by Eq.(\ref{kala}).
Substituting this into Eq.(\ref{master}), 
and expressing $Z_N$ in a power series in $x$, we obtain~\cite{grossman2} 
\begin{equation}
Z_N = x^{E_0}\sum_{k=1}^{\infty} \Omega(E_k^{(ex)},N) x^{E_k^{(ex)}},
\label{series} 
\end{equation}
where the $N$-particle eigen-energies $E_k=E_0+E_k^{(ex)}$ form an ordered 
set, with $E_0$ and $E_k^{(ex)}$ denoting the ground state energy and  
the excitation energy with respect to the ground state respectively. 
The expansion
coefficient $\Omega(E_k^{(ex)}, N) $ denotes the number of possible ways
of distributing the excitation energy $E_k^{(ex)}$ in utmost N
particles\cite{grossman1,grossman2}.

Furthermore we may write  $\Omega(E_k^{(ex)}, N)$ as
\begin{equation}
\Omega(E_k^{(ex)}, N) = \sum_{N_{ex} =1}^{N} \omega(E_k^{(ex)},N_{ex},N)  
\label{omega}
\end{equation} 
where $\omega(E_k^{(ex)},N_{ex}, N) $ denotes the number of possible ways
of distributing the excitation energy $E_k^{(ex)}$ amongst {\it{exactly}}  
$N_{ex}$ particles. Hence the probability of exciting exactly
$N_{ex}$ particles from an $N$-particle system at an excitation energy
$E_k^{(ex)}$ is given by  
\begin{equation}
p(E_k^{(ex)},N_{ex},N) =  \frac{\omega(E_k^{(ex)},N_{ex}, N).
}{\Omega(E_k^{(ex)}, N)}, ~~~ N_{ex} = 0,...,N.
\label{probability}
\end{equation} 
By definition this probability is properly normalised. Further the
probability has the following properties:
\begin{eqnarray}
p(0,N_{ex},N) &=&  \delta_{0N_{ex}},\\
p(E_k^{(ex)},N_{ex},N) &=& 0 ~~~ N_{ex} > N. 
\end{eqnarray} 

The number fluctuation in the ground state of the system may now be
defined in terms of the moments of the probability distribution given
above. We first define the
moments: 
\begin{eqnarray}
<N_{ex}> &=& \sum_{N_{ex}=1}^{N} N_{ex} p(E_k^{(ex)},N_{ex},N),\\
<N_{ex}^2> &=& \sum_{N_{ex}=1}^{N} N_{ex}^2 p(E_k^{(ex)},N_{ex},N),
\end{eqnarray}
and the number fluctuation from the ground state is given by,
\begin{eqnarray}
\delta N_0^2 &=& <N_{ex}^2> - <N_{ex}>^2~, \nonumber \\
             &=& <N_0^2> - <N_0>^2~,
\label{deltanzero}
\end{eqnarray}
since $<N_0>+<N_{ex}>=N$.

Few remarks are in order here: The above definitions apply equally well to
bosonic and fermionic systems. The fluctuation in the number 
of particles from the ground state is expressed here as a function of the 
excitation energy with respect to the ground state. In the case of bosons this
is just the fluctuation from the lowest energy single-particle state, 
where as for fermions it is
the number fluctuation across the (zero temperature) Fermi energy.  
Formally the above expressions
complete the necessary basic definitions for further analysis. 
In actual practice the counting of the number of possibilities to which a 
system may be excited is non-trivial. In the following, we give an example 
of bosons and fermions in a harmonic oscillator potential for the explicit 
calculation of $\omega(E_k^{(ex)}, N_{ex}, N)$. Henceforth, we consider  
the excitation energy from the ground state to be $n \hbar\omega$, and denote 
the corresponding microcanonical multiplicity to be 
$\omega(n, N_{ex}, N)$. \\

(a) Bosons in a $d$-dimensional harmonic oscillator.\\ 
In the case of a harmonic oscillator trap, all the single particle 
energy levels in a given shell have the same energy. The shell may be 
characterised by the index $k$ with energy $(k-1+d/2)\hbar\omega, ~
k=1,2,..\infty$, and degeneracy $g_k$. When the excitation energy is zero
(equivalently at zero temperature) all the $N$ particles are in the ground
state. Now consider exciting $N_{ex}$ particles from this ground state
that share $n$ quanta of energy. Let there be $m_{1}$ bosons with 
1 quantum of excitation energy, $m_{2}$ bosons with 2 quanta of excitation, 
and so on. We then have  
$$ n = m_1  + 2 m_2  + ... + (n+1 - N_{ex}) m_{N_{ex}}$$
and
$$ N_{ex} = m_1 +m_2 + ...+ m_{N_{ex}}.$$
This is just a way of partitioning an integer $n$ into exactly 
$N_{ex}$ partitions with $m_i$ denoting the number of times an integer 
$i$ appears in a partition ( it is taken to be zero if some 
integer between 1 and
$n+1-N_{ex}$ does not appear in the partition). All the $m_i$ bosons 
occupy the state $i+1$, since $i=1$ is the ground state. 
The number of ways these $m_i$ bosons are distributed in the state 
$i+1$ is given by the counting rule
\begin{equation}
{^{(g_{i+1}+m_i-1)}}C_{m_i}.
\label{late}
\end{equation}
The set $\{m_i\}$ denotes the set of all partitions of n for a given
$N_{ex}$. The microcanonical distribution $\omega$ is then given by 
\begin{equation}
\omega(n,N_{ex},N)=\sum_{\{m_k\}}\prod_{k=1}^{N_{ex}}
{^{(g_{k+1}+m_{k}-1)}}C_{m_{k}}.
\end{equation}
The formula for $\omega$ is especially simple in one dimension since there
is no degeneracy, $g_k=1$ for all $k$.
We have,
\begin{equation}
\omega(n,N_{ex},N)=\sum_{\{m_k\}} 1,
\end{equation}
and the number of possibilities are determined by the number of ways an
integer $n$ can be partitioned into $N_{ex}$ integers, that is the number
of ways we can write,
$$n= n_1+n_2+n_3+...+n_{N_{ex}}, ~~~ n_1 \ge n_2 \ge ...n_{N_{ex}}.$$
Once the $\omega$'s are known the probability distribution may be
calculated using the Eq.(\ref{probability}) and hence the fluctuations as
a function of the excitation energy. In order to express this as a
function of temperature, one may use the relation between excitation energy
above the ground state and the temperature to be discussed presently 
(see Eq.(\ref{extotemp}). We, however, prefer to present all results as a 
function of the excitation energy only.
Since $N_{ex} \le N$ always, we can give an alternative expression for
$\omega$ as~\cite{grossman2} 
\begin{equation}
\omega(n,N_{ex},N)=\Omega(n,N_{ex}) - \Omega(n,N_{ex}-1).
\label{omegasimple}
\end{equation}
While this allows an alternative and probably a simpler way of calculating
probability distribution, it is valid only for bosons 
(see the discussion following Eq.(\ref{mushkil}) given later).\\

(b) Fermions in a harmonic-oscillator (closed shell).\\
For simplicity, the fermions are taken to 
be spinless, and filling an integral number of shells in the ground state.
The determination of microcanonical distribution $\omega(n,N_{ex},N)$ is 
more
involved than in the case of bosons. The combinatoric rules depend on 
the space dimension. Using the same notation for the harmonic 
oscillator single particle energies in a given shell,  
we calculate the number of possibilities by which $N_{ex}$
particles can be excited from the ground state.  Note that removing
$N_{ex}$ fermions from the ground state leaves as many holes in ground
state. As a result the $\omega$ depends not only on the distribution of
fermions in the excited states (as in the case of bosons), but also on how
the holes are distributed in the ground state. 
For simplicity we choose a
completely closed shell system, the rules may be appropriately modified
for open shells. We assume that N-fermions form a closed-shell 
system at $T=0$ with shells up to $k$ (corresponding to 
fermi energy $E_F=(k-1+d/2)\hbar\omega$) filled.
Thus $N_{ex}$ holes
are distributed in states from $i=1$ up to $i=k$. Let $h_i$ denote the
number of holes in the state $i$ such that 
$$N_{ex} = h_1 + h_2 + ...+h_k.$$
The number of ways $h_i$ holes may be
created in the $i$th state is given by ${^{g_i}}C_{h_i}$. 
Hence the number of ways in which $N_{ex}$ holes may be created in 
the ground state is given by the product over the number of ways in 
which the holes may be distributed in the $k$ orbitals, i.e., 
$$\prod_{j=1}^{k}{^{(g_{j})}}C_{h_{j}}.$$  
Now consider exciting $N_{ex}$ particles from this ground state sharing
$n$ quanta of energy. An allowed configuration is one in which each and
every one of these $N_{ex}$ particles is found in states above the 
fermi energy, with the   
shell indices ranging from $(k+1)$  up to $(k+n)$, such that their 
excitation energies add up to yield the total $E_{ex}$. This 
complicates the
counting rules for fermions as compared to bosons. We shall denote the
occupancy of orbitals for the excited particles by $m_i$, where
$i=k+1,...,k+n$. The number of ways the $m_i$ femions are distributed 
in the state $k+i$ is then given by the counting rule
$${^{(g_{k+i})}}C_{m_i}.$$
The microcanonical distribution $\omega$ is then given by
\begin{equation}
\omega(n,N_{ex},N)=
\sum_{\{m_i\}}\sum_{\{h_j\}}
\prod_{j=1}^{k}{^{(g_{j})}}C_{h_{j}}
\prod_{i=k+1}^{k+n}{^{(g_{k+i})}}C_{m_{i}},
\end{equation}
where $N_{ex}=\sum_i~m_i$, and 
the microcanonical multiplicity $\omega$ is obtained by summing over all the
allowed possibilities such that the sum total of the excited quanta 
is exactly $n$.
Once the $\omega$'s are known the probability distribution may be
calculated using the Eq.(\ref{probability}) and hence the fluctuation as
a function of the excitation energy. 

We compare these results with the fluctuations obtained
by the canonical ensemble averaging method of 
Parvan {\it et al.}~\cite{parvan} as detailed below. It is our objective to 
see how close are the results for the ground state fluctuations calculated 
by the two methods. 

\subsection{Fluctuations from canonical ensemble averaging}
In statistical thermodynamics, a macro-state at a given energy may be 
built up by many combinatorial ways from the micro-states, and 
this is denoted by the term multiplicity of the macro-state. 
As we saw in the preceding sect.(II A), the multiplicity 
$\Omega(E_k^{ex},N)$ was defined by Eq.(\ref{omega}) through this 
counting method. In the canonical ensemble, we may alternately 
define 
\begin{equation}
\Omega(T,N)=\exp[S(T,N)]~=\exp[(U-F)/T]~=x^{-U} Z_N(x)~. 
\label{thermalomega}
\end{equation} 
In the above, the internal energy $U(T)$ is determined as usual from the 
canonical partition function, and excitation energy at any
temperature is given by 
\begin{equation} E^{(ex)} = U(T) - U(0)~. 
\label{extotemp} 
\end{equation} 
We may therefore compare the calculated quantities from the canonical and 
the microcanonical ensembles as a function of the excitation energy.

Comparing $\Omega(T,N)$ given by Eq.(\ref{thermalomega}) from 
the canonical ensemble 
with the  series (\ref{series}), we see that it is as if only one 
term from this series is picked in the ensemble averaging. This is  
realized for a large number of particles, since the multiplicity 
$\Omega(E_k^{ex},N)$ increases rapidly with the excitation energy, 
where as the factor $x^{E_k}$ decreases exponentially. In this 
paper, we focus on systems where $N$ is not large, specially for 
fermions. It is therefore interesting to examine the differences 
in the results of the calculation made by the two methods.

Once the multiplicity $\Omega$ is obtained, the fluctuations may 
be defined through a temperature dependent
probability distribution\cite{grossman1,grossman2},
\begin{equation}
p(T,N_{ex},N) = \frac{\Omega(T,N_{ex}) - \Omega(T,N_{ex}-1)}{\Omega(T,N)},
~~~N_x=0,...,N .
\label{mushkil}
\end{equation}
Note that $\Omega(T,N_{ex})$ is determined by the canonical partition
function of $N_{ex}$ particles by replacing $N$ by $N_{ex}$ in
Eq.(\ref{thermalomega}). 
For bosons, the ground state of every such system consists of a single 
level, and the definition given by Eq.(\ref{mushkil}) is correct. 
For fermions, however, the Fermi energy of a system defined by 
$N_{ex}$ particles is less than the Fermi energy of the full system
with $N$ particles. The fluctuations are defined with respect to the 
ground state with the Fermi energy corresponding to the full system. 
Therefore Eq.(\ref{mushkil}) cannot be applied to fermions. 
A consistent definition of
ground state fluctuations applicable to both fermions and bosons, and which 
coincides for the bosonic calculation with the probability definition 
(\ref{mushkil}), may however be given by the ensemble averaging 
method~\cite{zamick,parvan}. We summarise the method below.  

The canonical partition function in 
Eq.(\ref{series}) may be written in the occupation number 
representation as\cite{parvan}
\begin{equation}
Z_N = \sum_{\{n_k\}}\prod_{k} x^{\epsilon_k n_k},
\label{occnumrep} 
\end{equation}
where we have used the fact that the energy of the $N$- particle system
for a given set of occupancies $\{n_k\}$ is given by $E = \sum_k
\epsilon_k n_k$. The occupancy $n_k=0,1$ for fermions and may take any 
value up to $N$ for bosons. At finite temperatures,  
using the recursion relation in Eq.(\ref{recursion}), and some nontrivial 
algebra, the ensemble-averaged moments of the occupancy $n_k$ may be  
written as~\cite{parvan}
\begin{eqnarray}
<n_k> &=& \frac{1}{Z_N} \sum_{j=1}^N
(\pm)^{j+1} x^{j\epsilon_k} Z_{N-j} ,
\label{avnkx}\\
<n_k^2> &=& \frac{1}{Z_N} \sum_{j=1}^{N} (\pm)^{j+1}j
x^{j\epsilon_k}Z_{N-j}
\nonumber\\
&+& \frac{1}{Z_N} \sum_{j=1}^{N}\sum_{i=1}^{N-j}(\pm)^{i+j}
x^{(i+j)\epsilon_k}Z_{N-i-j},
\label{avnksx}
\end{eqnarray}
where the upper and lower signs refer to bosons and fermions respectively.

The ground state fluctuation is now given by,
\begin{equation}
\delta N_0^2 = \sum_{k}(<n_k^2> - <n_k>^2).
\label{thermalfluct}
\end{equation}
The sum runs through all the allowed $k$ values in the ground state
defined at zero temperature. Note that the moments of occupancy are
defined without any reference to the ground state. The fluctuation is then
obtained by summing over the single particle states occupied at zero
temperature. In the counting method as outlined earlier, the fluctuations
necessarily refer to the ground state. 

Applying the above equations to a system of bosons in a harmonic trap, we
get
\begin{eqnarray}
<n_0>& = &\frac{1}{Z_N} \sum_{j=1}^N x^{jd/2} Z_{N-j} ,
\label{bavnkx}\\
<n_0^2> &=& \frac{1}{Z_N} \sum_{j=1}^{N} (2j-1)x^{jd/2}Z_{N-j}.
\label{bavnksx}
\end{eqnarray}
The ground state fluctuation is given by Eq.(\ref{thermalfluct}),
\begin{equation}
\delta N_0^2 = (<n_0^2> - <n_0>^2).
\end{equation}

Similarly, the thermodynamic expression for the ground state fluctuations for 
the fermionic 
system in a harmonic oscillator is obtained from Eq.(\ref{avnkx}) and 
Eq.(\ref{avnksx}):
\begin{eqnarray}
<n_s>& = &\frac{1}{Z_N} \sum_{j=1}^N (-1)^{j+1} g_s x^{j(s-1+d/2)} Z_{N-j} ,
\label{favnkx}\\
<n_s^2> &=& \frac{1}{Z_N}
\sum_{j=1}^{N}(-1)^{j+1}(j g_s-(j-1)g_s^2)x^{j(s-1+d/2)}Z_{N-j},
\label{favnksx}
\end{eqnarray}
where $g_s$ refers to the degeneracy factor of the the state $s$. 
For the N-fermion system that forms a closed shell system at $T=0$ with 
filled shells up to $k$, with Fermi energy $E_F=(k-1+d/2)\hbar\omega$, the 
ground state fluctuation is given by Eq.(\ref{thermalfluct}):
\begin{equation}
\delta N_0^2 = \sum_{s=1}^k~(<n_s^2> - <n_s>^2).
\label{fence}
\end{equation}

Before using these formulae for the calculation of ground state occupancy 
and fluctuations, we discuss below the specially interesting case of 
the one-dimensional harmonic confinement where the canonical and the 
microcanonical ensembles yield vastly different results.

\section{ Fluctuations in a one-dimensional harmonic trap}
The one-dimensional Harmonic trap is specially interesting because 
even though the canonical entropies for bosons and fermions are
identical, the number fluctuations from the ground state are very 
different. This may be seen by writing the canonical $N$-particle 
partition function as 
\begin{equation}
Z_N = x^{gN(N-1)/2 + N/2} \prod_{j=1}^N \frac{1}{(1-x^j)},
\label{chacha}
\end{equation}
with $g=0$ for bosons, and $g=1$ for fermions. Actually, for other
positive values of $g$, the above form is the exact partition function 
for the so-called Calogero-Sutherland model~\cite{CSM}, where the 
$N$ particles interact pair-wise by a potential 
${\hbar^2\over m}\sum_{i<j}^N (x_i-x_j)^{-2}$. For bosons, the
dimensionless parameter $g$ is in the range $0\leq g\leq 1/2$, while
for fermions $g > 1/2$. The special values $g=1(0)$ give
noninteracting fermions (bosons). 
The effect of interaction has only been to shift the energy of every state 
by the same amount, which is absorbed in the prefactor. It follows from 
Eqs.(\ref{series},\ref{chacha}) that we may write 
\begin{equation}
Z_N = x^{gN(N-1)/2 + N/2} \sum_{n=0}^{\infty} \Omega(n,N) x^n~,
\end{equation}
and that $\Omega(n,N)$ is independent of the parameter $g$, and 
is the same for bosons and fermions. Since it is the logarithm 
of $\Omega(n,N)$ that determines the canonical entropy of the system 
at an excitation energy of $n$ quanta, it follows that the entropy is 
independent of $g$. The same result is true if one calculates the
ensemble averaged entropy at a given temperature. This may be easily 
verified by using the relation $F=- \ln Z_N/\beta$ for the free energy, 
and then calculating the entropy $S=-{\partial F\over {\partial T}}$.  

For the microcanonical calculation of the fluctuation, we need to calculate 
the microcanonical multiplicity $\omega(n,N_{ex},N)$, which is the number of 
ways of distributing the $n$ excitation quanta amongst exactly $N_{ex}$ 
particles. Although the relation 
\begin{equation}
\Omega(n,N)=\sum_{N_{ex}=1}^N~\omega(n,N_{ex},N)
\end{equation}
is obeyed both by fermions and bosons, and the LHS of the above equation is 
the same for both, the microcanonical counting of $\omega$'s are very 
different for the two cases. In the appendix, this is illustrated  
explicitly for $N=3$.  
Thus the fluctuations for the bosonic and fermionic cases differ 
substantially when the exact counting method is used. As is to be 
expected, the fermionic fluctuation is considerably suppressed
compared to the bosonic system. The same qualitative conclusion may also be
reached by using Eqs.(\ref{bavnksx}, \ref{favnksx}) based on  
canonical ensemble averaging rather than exact counting.  
In the next section, we display these results, as well as the results for 
the two-dimensional harmonic oscillator. 

\section{ Results and Discussion}
In Fig.(1), we display (a) the ground state occupancy $<N_0>$ and 
(b) the relative fluctuation $<\delta N_0>/N$ for $N=15$ 
noninteracting particles in 
a one-dimensional harmonic oscillator potential. The canonical and the 
exact microcanonical results are compared as a function of the excitation 
energy. The canonical method gives results in close agreement with
counting for the ground state occupancy $<N_0>$, but overestimates the 
relative fluctuation substantially.  
In one dimension, the number of microcanonical possibilities 
$\omega(n,N_{ex},N)$  
is very restricted at low excitations due to the non- degeneracy in the 
single-particle energy levels and the Pauli exclusion principle. 
This results in (i)$<N_0>$ for fermions getting depleted 
more slowly than bosons, and (ii) reduced fluctuation.  
In Fig.(2), the same quantities are displayed for particles in a 
two-dimensional harmonic oscillator potential. It is well known from 
previous work\cite{grossman1,grossman2,grossman3} that there is a peak in 
the relative fluctuation
somewhat below the critical temperature for bosons. For fermions, such 
peaking is absent, specially in the exact calculation. We thus see
that the presence or absence of a peak in the number fluctuation of
the ground state may signal phase transition, or its absence.
 
The microcanonical method of exact counting for fermions is
computationally very time-consuming because, unlike bosons,
Eq.(\ref{mushkil}) cannot be used. We have therefore restricted the
fermionic calculations to only up to $N=15$ particles. As we see from Figs
(1-2), there is considerable difference in the results for the relative
fluctuations for small particle numbers. For fermions, we expect this
difference to persist at low excitations even when $N$ is large. This is
because at low excitations, only a small fraction of the fermions near the
Fermi sea can be excited, so the effective number of fermions contributing
to the number fluctuation remains small even when the system is large. For
bosons, there is no difficulty in performing the microcanonical
calculation for a large number of particles. In Fig.(3), we show the
bosonic occupancy and the relative fluctuation as a function of the number
of excitation quanta for $N=100$. We note that the occupancy $<N_0>$ from
the canonical and exact microcanonical methods is practically identical,
although the ground state fluctuation continue to differ. 

We have made the comparison between the microcanonical and canonical results 
as functions of the excitation energy, rather than temperature. In the 
microcanonical method, the counting of the possibilities is done for a 
given excitation energy, the latter being the natural variable. In the 
canonical formalism, of course, a mapping from the excitation energy to 
temperature can be done using Eq.(\ref{extotemp}). 
We may also note that starting from the canonical Eq.(\ref{thermalomega}) and 
using Eq.(\ref{mushkil}) for bosons, one can obtain the analytical 
expression for the low temperature dependence of the fluctuation, 
as, for example, given by Eq.(14) of \cite{grossman2}. It remains a 
challenging problem to obtain similar microcanonical relations for 
fermions.  

This research was supported by the Natural Sciences and Engineering Research 
Council (NSERC) of Canada. M.T. would like to acknowledge receiving an 
NSERC post-graduate scholarship, and M.V.N.M. enjoyed the hospitality of 
the department of Physics and Astronomy of Mcmaster University, where the 
work was done. 
\newpage
\begin{figure} 
\label{figure1}
\caption{(a) Plots of the ground state occupancy $<N_0>$ versus the 
excitation energy $E_{ex}$ for $N=15$ bosons (fermions) in a one-dimensional 
harmonic oscillator. $E_{ex}$ is given in units of the oscillator spacing.
The microcanonical results 
are displayed according to the legends in the inset box.
(b) Plots of the relative ground state fluctuation $<\delta N_0>/N$ 
as a function of $E_{ex}$ for the same systems as in (a).} 
\vspace{7 mm}
\label{figure2}
\caption{(a) Plots of the ground state occupancy $<N_0>$ versus the 
excitation energy $E_{ex}$ for $N=10$ bosons (fermions) in a two-dimensional 
harmonic oscillator, according to the legends in the inset box.
(b) Plots of the relative ground state fluctuation $<\delta N_0>/N$ 
as a function of $E_{ex}$ for the same systems as in (a).} 
\vspace{7 mm}
\label{figure3}
\caption{(a) Plots of the ground state occupancy $<N_0>$ versus the 
excitation energy $E_{ex}$ for $N=100$ bosons in a one-dimensional 
harmonic oscillator. (b) Plots of the relative ground state 
fluctuation $<\delta N_0>/N$ 
as a function of $E_{ex}$ for the same system as in (a).} 
\end{figure}
\newpage
\section{Appendix}
Calculation of $\omega(n,N_{ex},N)$ in one-dimensional harmonic oscillator.\\
(a) {\bf Bosons}\\
For illustration, in Table 1, we display the  simple case of $N=3$ 
bosons, with the the number of excitation quanta $n\leq 7$. Since there 
is no degeneracy, the combinatorial factor given by (\ref{late}) 
is unity. Therefore $\omega(n,N_{ex},N)$ in this case is just the number 
of distinct ways in which the integer $n$ may be partitioned amongst 
exactly $N_{ex}$ identical bosons $(N_{ex}\leq 3)$. In Table 1, 
$n$ increases from left to right, and increasing values of $N_{ex}$ 
are tabulated in a vertical column. The integer in each box is the 
corresponding  $\omega(n,N_{ex},N)$, with the distinct partitions of 
$n$ listed in brackets below it. For 
example, in the box under $(n=4, N_{ex}=2)$, we see that 
$\omega(4,2,3)=2$, and the two distinct partitions of $4$ are $(3+1)$ 
and $(2+2)$. In the last row is listed $\Omega(n,N)$, that is obtained 
by adding the $\omega(n,N_{ex},N)$'s in each vertical column. Note 
that these check with the coefficients in the expansion of the $3$-boson
canonical partition function :\\
$$Z_3=1+x+2 x^2 +3 x^3 + 4 x^4 + 5 x^5 +7 x^6 + 8 x^7 +...  .$$   

(b) {\bf Fermions}\\
Same as for bosons, except that two more vertical columns have been added 
under the headings $L_1$ and $n_{min}$. For three spinless fermions, the 
lowest three energy levels $(1,2,3)$ (in increasing order of energy) are 
occupied at $T=0$. The column 
under $L_1$ lists the possible configuration of holes in these levels 
for an excitation 
energy of $n\geq n_{min}$ quanta, the latter being listed in the adjacent 
column. For example, for $N_{ex}=2$, $(3_12_1)$ under the column $L_1$ 
denotes a two-hole configuration, with one hole in level  $3$, and 
another in level $2$. The minimum energy required for this is $n_{min}=4$ 
in units of $\hbar\omega$.

\begin{table}
\begin{center}
\caption[table1]{Tabulation of bosonic $\omega(n,N_{ex},N)$ for N=3}

\begin{tabular}{|c c|c|c|c|c|c|c|c|}
\multicolumn{1}{|c}{} &n=  &1 &2 &3 &4 &5 &6 &7    \\ \hline
\multicolumn{1}{|c}{} &$N_{ex}$=1 &1 &1 &1 &1 &1 &1 &1 \\ \cline{2-9}
$\omega(n,N_{ex},N)$:\rule{0in}{.70in} &$N_{ex}$=2 &\shortstack{0\\ \rule{0in}{0.5in}} &\shortstack{1\\(1+1)\\ \rule{0in}{0.3in}} &\shortstack{1\\(2+1)\\ \rule{0in}{0.3in}} 
&\shortstack{2\\(3+1)\\(2+2)\\ \rule{0in}{0.1in}} &\shortstack{2\\(4+1)\\(3+2)\\ \rule{0in}{0.1in}} 
&\shortstack{3\\(5+1)\\(4+2)\\(3+3)} &\shortstack{3\\(6+1)\\(5+2)\\(4+3)} \\ \cline{2-9}
\multicolumn{1}{|c}{\rule{0in}{.90in}} &$N_{ex}$=3 &\shortstack{0\\ \rule{0in}{0.7in}} &\shortstack{0\\ \rule{0in}{0.7in}} &\shortstack{1\\(1+1+1)\\ \rule{0in}{0.5in}} &\shortstack{1\\(2+1+1)\\ \rule{0in}{0.5in}} 
&\shortstack{2\\(3+1+1)\\(2+2+1)\\ \rule{0in}{0.3in}} &\shortstack{3\\(4+1+1)\\(3+2+1)\\(2+2+2)\\ \rule{0in}{0.1in}}
&\shortstack{4\\(5+1+1)\\(4+2+1)\\(3+3+1)\\(3+2+2)} \\ \hline
$\Omega(n,N)=$ &\multicolumn{1}{c|}{} &1 &2 &3 &4 &5 &7 &8                   
\end{tabular}
\end{center}
\end{table}

\begin{table}
\begin{center}
\caption[table2]{Tabulation of fermionic $\omega(n,N_{ex},N)$ for N=3}
\begin{tabular}{|c c|c|c|c|c|c|c|c|c|} 
\multicolumn{3}{c|}{} &\multicolumn{7}{c}{n} \\ \hline 
\multicolumn{1}{|c}{} &$L_{1}$ &$n_{min}$ &1 &2 &3 &4 &5 &6 &7  \\ \hline \hline 
\multicolumn{1}{|c}{} &$3_{1}$ &1 &1 &1 &1 &1 &1 &1 &1          \\ \cline{2-10}
$N_{ex}=1:$ &$2_{1}$ &2 &0 &1 &1 &1 &1 &1 &1  \\ \cline{2-10}
\multicolumn{1}{|c}{} &$1_{1}$ &3 &0 &0 &1 &1 &1 &1 &1  \\ \cline{2-10}
\multicolumn{1}{|c}{$\omega(n,1,N)=$} &\multicolumn{2}{c|}{} &1 &2 &3 &3 &3 &3 &3 \\ \hline 
\multicolumn{1}{|c}{\rule{0in}{0.5in}} &$3_{1}2_{1}$ &\shortstack{4\\ \rule{0in}{0.3in}} &\shortstack{0\\ \rule{0in}{0.3in}}  &\shortstack{0\\ \rule{0in}{0.3in}} &\shortstack{0\\ \rule{0in}{0.3in}} &\shortstack{1\\(1+3)\\ \rule{0in}{0.1in}} &\shortstack{1\\(1+4)\\ \rule{0in}{0.1in}} &\shortstack{2\\(1+5)\\(2+4)} &\shortstack{2\\(1+6)\\(2+5)} \\ \cline{2-10}  
$N_{ex}=2:$\rule{0in}{0.5in} &$3_{1}1_{1}$ &\shortstack{5\\ \rule{0in}{0.3in}} &\shortstack{0\\ \rule{0in}{0.3in}} &\shortstack{0\\ \rule{0in}{0.3in}} &\shortstack{0\\ \rule{0in}{0.3in}} & \shortstack{0\\ \rule{0in}{0.3in}}&\shortstack{1\\(1+4)\\ \rule{0in}{0.1in}} &\shortstack{1\\(1+5)\\ \rule{0in}{0.1in}} &\shortstack{2\\(1+6)\\(2+5)} \\ \cline{2-10}
\multicolumn{1}{|c}{\rule{0in}{0.3in}} &$2_{1}1_{1}$ &\shortstack{6\\ \rule{0in}{0.1in}} &\shortstack{0\\ \rule{0in}{0.1in}} &\shortstack{0\\ \rule{0in}{0.1in}} &\shortstack{0\\ \rule{0in}{0.1in}} &\shortstack{0\\ \rule{0in}{0.1in}} &\shortstack{0\\ \rule{0in}{0.1in}} &\shortstack{1\\(2+4)} &\shortstack{1\\(2+5)}  \\ \cline{2-10}
\multicolumn{1}{|c}{$\omega(n,2,N)=$} &\multicolumn{2}{c|}{} &0 &0 &0 &1 &2 &4 &5  \\ \hline 
$N_{ex}=3:$ &$3_{1}2_{1}1_{1}$ &9 &0 &0 &0 &0 &0 &0 &0 \\ \cline{2-10}
\multicolumn{1}{|c}{$\omega(n,3,N)=$} &\multicolumn{2}{c|}{} &0 &0 &0 &0 &0 &0 &0  \\ \hline 
\multicolumn{1}{|c}{$\Omega(n,N)=$} &\multicolumn{2}{c|}{} &1 &2 &3 &4 &5 &7 &8\\          
\end {tabular}
\end{center}
\end{table}

\end{document}